\documentclass[a4paper,11pt]{article}
\usepackage{pos}
\usepackage{here}
\usepackage{subcaption}

\title{Single-charged Higgs boson in $W^\pm H^\mp$ associated production within the 2HDMs.}
\author[a]{A. Arhrib}
\author[b]{R. Benbrik}
\author[c]{M. Krab}
\author[c]{B. Manaut}
\author*[b]{M. Ouchemhou}
\author[d,e]{Qi-Shu Yan}

\affiliation[a]{Abdelmalek Essaadi University, Faculty of Sciences and techniques, Tanger, Morocco.}

\affiliation[b]{Polydisciplinary Faculty, Laboratory of Fundamental and Applied Physics, Cadi Ayyad University,\\ Sidi Bouzid, B.P. 4162, Safi, Morocco.}

\affiliation[c]{Research Laboratory in Physics and Engineering Sciences, Modern and Applied Physics Team, Polydisciplinary Faculty, Beni Mellal, 23000, Morocco.}

\affiliation[d]{Center for Future High Energy Physics, Chinese Academy of Sciences, Beijing 100049, PR China.}

\affiliation[e]{School of Physics Sciences, University of Chinese Academy of Sciences, Beijing 100039, PR China.}

\emailAdd{aarhrib@gmail.com}
\emailAdd{r.benbrik@uca.ac.ma}
\emailAdd{mohamed.krab@usms.ac.ma}
\emailAdd{b.manaut@usms.ma}
\emailAdd{mohamed.ouchemhou@ced.uca.ac.ma}
\emailAdd{yanqishu@ucas.ac.cn}

\abstract{In this contribution, the likelihood of seeing charged Higgs and $W$ boson production in the context of 2HDMs type-I and type-X at the LHC is examined, assuming
that either $h$ or $H$ resembles the detected resonance around $\sim 125$ GeV. We consider the possibility of the charged Higgs boson decays channels through $H^\pm \to W^\pm h_i / A$, focusing on the $b\bar{b}$ and $\tau\tau$ decays of $h_i$ and $A$. In both type-I and type-X insights of the 2HDMs, we investigate the potential fingerprints resulting from the previously mentioned charged Higgs production and decay. We find in our study that these signatures can have sizable rates at low $\tan\beta$ as long as the condition $M_{H^\pm} < m_t - m_b$ is met. As a result, we propose the $bb$ and $\tau\tau$ final states associated with $WW$ as an encouraging experimental avenue that would complement the LHC search for a charged Higgs boson.}

\FullConference{%
  8th Symposium on Prospects in the Physics of Discrete Symmetries (DISCRETE 2022)\\
  7-11 November, 2022\\
  Baden-Baden, Germany
}

\begin{document}
\maketitle
\section{Introduction}
\label{sec:intro}
Through experimental accuracy tests over the past few years, the Standard Model of particle physics has achieved a remarkable degree of success. Despite its popularity, the key components of the theory—the Higgs mechanism and the issue of the mass spectrum—remain unanswered and out of reach of the most current research. The Higgs sector issues appear to have appealing candidate answers in theoretical frameworks outside of the SM. The Two-Higgs Doublet model with a cp-conserving extension to the SM, the so called 2HDMs\cite{Branco:2011iw}, appears as a simple, reliable and minimal model, which predicts five physical Higgs bosons. Two scalar $h$ and $H$ which may be recognized as the state discovered around $126$ GeV by the LHC \cite{ATLAS:2012yve,CMS:2012qbp}, a pseudo-scalar $A$ and two charged Higgs bosons $H^\pm$. The observation of a charged Higgs is a crucial signature of the model beyond the SM. This is a sufficient justification for why this particle has drawn special interest in recent years in many high-energy physics experiments. The LHC searches for the charged Higgs boson through many distinct production modes, taking into account the lightness $m_{H^\pm} \leq m_t-m_b$ or the heavyness $m_{H^\pm} \geq m_t-m_b$ limit. The former limit allowed a production from the top pair production $t\bar{t}$ while the last allowed the production through $gg \to tbH^-$ and $gb \to tH^-$. In this work, we review the production of the charged Higgs boson along with the $W$ boson within the 2HDMs type-I and type-X frameworks\cite{Arhrib:2022ehv}. These models still predict a light-charged Higgs boson with significant rates of its bosonic decays that could potentially dominate over fermionic ones. In the context of the LHC, we investigate the various potential signatures resulting from the previously mentioned Higgs production channels and the bosonic decays $H^\pm \to W^\pm h_i / A$, as well as their phenomenological Implications. This contribution is presented as follows: We review the 2HDMs framework in Section\ref{sec:model}. The results of our research will be discussed in section\ref{sec:Results}. The benchmark points (BPs) are set in Section \ref{sec:BPs}, while section \ref{sec:conclusion} is devoted to the conclusion.

\section{The 2HDMs framework}
\label{sec:model}
Under the softly broken $Z_2$ symmetry, the 2HDM's scalar potential, which is $SU(2)_L\otimes U(1)_Y$ invariant and CP-conserving, is given by:
\begin{align}
V_{\rm{2HDM}} ~&=~ m_{11}^2(\Phi_1^+\Phi_1)+m_{22}^2(\Phi_2^+\Phi_2)-m_{12}^2(\Phi_1^+\Phi_2+ \rm{h.c.})+\lambda_1(\Phi_1^+\Phi_1)^2+\lambda_2(\Phi_2^+\Phi_2)^2\nonumber \\
&~ +\lambda_3(\Phi_1^+\Phi_1)(\Phi_2^+\Phi_2)
+\lambda_4 (\Phi_1^+\Phi_2)(\Phi_2^+\Phi_1)+\frac{\lambda_5}{2}[(\Phi_1^+\Phi_2)^2+ \rm{h.c.}], \label{RTHDMpot}
\end{align}
The $m_{11},m_{22}$ and $m_{12}$ are masses parameters whereas the $\lambda_i$ are couplings parameter with no dimensions, they are all real due to the cp-conserving proprieties. The doublets $\Phi_1$ and $\Phi_2$ gains vacuum expectation values $v_1$ and $v_2$ via Spontaneous Electro-Weak Symmetry Breaking. The model left then with ten real parameters that could be reduced using the two minimization conditions of the potential, treading $m_{11}$ and $m_{22}$ by $v_{1,2}$. The model could then be fully described with the seven
free real, independent parameters, which are: the Higgs sector masses $M_h , M_H , M_A , M_{H^\pm}$, the mixing angles from the group eigenstate to the mass eigenstate in the charged Higgs and the CP-odd (CP-even) Higgs sector $ \alpha, \tan\beta(=\frac{v_2}{v_1})$, and the $Z_2$ soft broken parameter $m_{12}^2$. 

The suppression of the FCNCs was done by introducing a $Z_2$ symmetry, which led to four versions of the model denoted, namely: type-I, type-II, type-X, and type-Y, respectively. However, due to the $B \to X_s \gamma$ constraints \cite{Haller:2018nnx}, a light-charged Higgs boson with $m_{H^\pm} \leq 680$  GeV is not yet survival in the type-II and type-Y models, as a results, our study is performed in the models type-I and X. The interactions between the fermions and Higgs sector are described by the Yaukawa Lagrangian, which takes the following form in the mass eigenstate basis:
\begin{align}
- {\mathcal{L}}_{\rm Yukawa} = \sum_{f=u,d,l} \left(\frac{m_f}{v} \xi_f^h \bar{f} f h + 
\frac{m_f}{v}\xi_f^H \bar{f} f H 
- i \frac{m_f}{v} \xi_f^A \bar{f} \gamma_5 f A \right) + \nonumber \\
\left(\frac{V_{ud}}{\sqrt{2} v} \bar{u} (m_u \xi_u^A P_L +
m_d \xi_d^A P_R) d H^+ + \frac{ m_l \xi_l^A}{\sqrt{2} v} \bar{\nu}_L l_R H^+ + \rm{h.c.} \right),
\label{Yukawa-1}
\end{align}
where $V_{ud}$ represents a CKM matrix element. the coefficients $\xi_f^{h_i, A}$ are 2HDM Higgs couplings to fermions normalized to the SM couplings, which are listed in Table \ref{coupling}.
\begin{table}[H]
	\begin{center} 
		\begin{tabular}{|c||c|c|c|c|c|c|c|c|c|} \hline
			&$\xi_{u}^{h}$&$\xi_{d}^{h}$&$\xi_{l}^{h}$&$\xi_{u}^{H}$&$\xi_{d}^{H}$&$\xi_{l}^{H}$&$\xi_{u}^{A}$&$\xi_{d}^{A}$&$\xi_{l}^{A}$\\\hline
			type-I & $c_\alpha/s_\beta$ & $c_\alpha/s_\beta$& $c_\alpha/s_\beta$ & $s_\alpha/s_\beta$ & $s_\alpha/s_\beta$ & $s_\alpha/s_\beta$ & $c_\beta/s_\beta$ & 
			$-c_\beta/s_\beta$ & $-c_\beta/s_\beta$ \\ \hline
			type-X & $c_\alpha/s_\beta$ & $c_\alpha/s_\beta$& $-s_\alpha/c_\beta$ & $s_\alpha/s_\beta$ & $s_\alpha/s_\beta$ & $c_\alpha/c_\beta$ & $c_\beta/s_\beta$ & 
			$-c_\beta/s_\beta$ & $s_\beta/c_\beta$ \\ \hline
		\end{tabular}
	\end{center}
	\caption{\label{coupling}Yukawa couplings of 2HDMs Higgs bosons to the fermions.}
\end{table}
\section{Results and discussion}
\label{sec:Results}
\begin{figure}[H]
	\centering
	\begin{tabular}{cccc}
		\begin{subfigure}[t]{.2\textwidth}\centering
			\includegraphics[height=1.5cm,width=2.5cm]{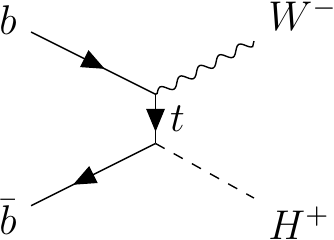}
			\caption{}
			\label{fig:Hptb_prod_bb_nonres}
		\end{subfigure}
		\hfill
		&\begin{subfigure}[t]{.2\textwidth}\centering
			\includegraphics[height=1.5cm,width=2.5cm]{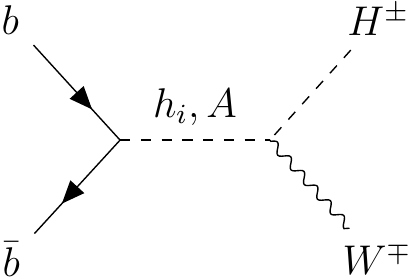}
			\caption{}
			\label{fig:Hptb_prod_bb_res}
		\end{subfigure}
		
		&\begin{subfigure}[t]{.2\textwidth}\centering
			\includegraphics[height=1.5cm,width=2.5cm]{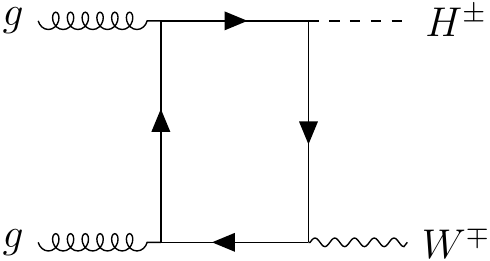}
			\caption{}
			\label{fig:Hptb_prod_gg_nonres}
		\end{subfigure}
		\hfill
		&\begin{subfigure}[t]{.2\textwidth}\centering
			\includegraphics[height=1.5cm,width=2.5cm]{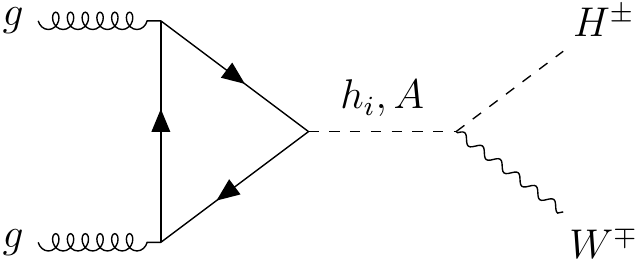}
			\caption{}
			\label{fig:Hptb_prod_gg_res}
		\end{subfigure}
	\end{tabular}
	\caption{Feynman diagrams contributing to the $H^\pm W^\mp$ production.}
	\label{fig:Hptb_prod_feynman}
\end{figure}

The four distinct subprocesses that make up the $pp \to H^\pm W^\mp$ production cross-section are shown by independent Feynman diagrams, as shown in Fig.\ref{fig:Hptb_prod_feynman}. The $b\bar{b}$ non-resonant and resonant channels are depicted in Figs. \ref{fig:Hptb_prod_bb_nonres} and \ref{fig:Hptb_prod_bb_res}, while the $gg$ initiated non-resonant and resonant channels are depicted in Figs. \ref{fig:Hptb_prod_gg_nonres} and  \ref{fig:Hptb_prod_gg_res}, respectively.

The free 2HDMs's parameters are randomly varied as set in table \ref{param_scans}, considering both scenario of H-like Higgs boson, Normal scenario(NS) and Inverted scenario (IS), and taking into accounts the full current available theoretical and experimental constraint. On the theoretical front, the vacuum stability, perturbativity, and unitarity pertubatives are forced, while on the experimental front, the collider bounds, Higgs signal strength, EWPOs through the oblique parameters ($S$, $T$, setting $U \neq 0$), and B-physics observables are also forced. These constraints are calculated by using the following public tools: 2HDMC\cite{Eriksson:2009ws}, HiggsBounds\cite{Bechtle:2020pkv}, HiggsSignal\cite{Bechtle:2020uwn}, SuperIso\cite{Mahmoudi:2008tp}.

\begin{figure}[H]
	\centering
	\begin{tabular}{cc}
		\includegraphics[width=0.46\textwidth]{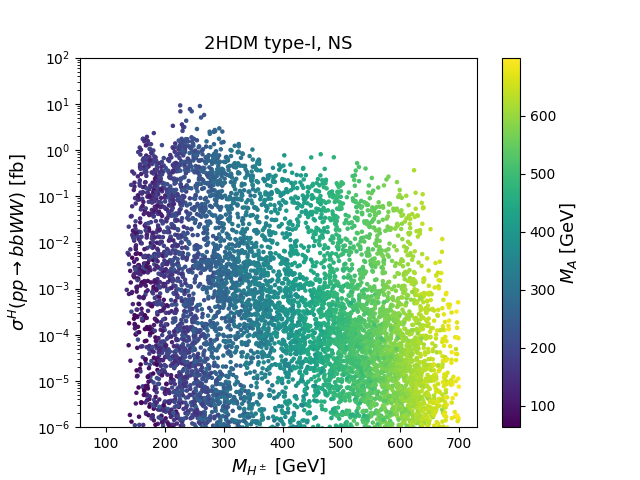} &
		\includegraphics[width=0.46\textwidth]{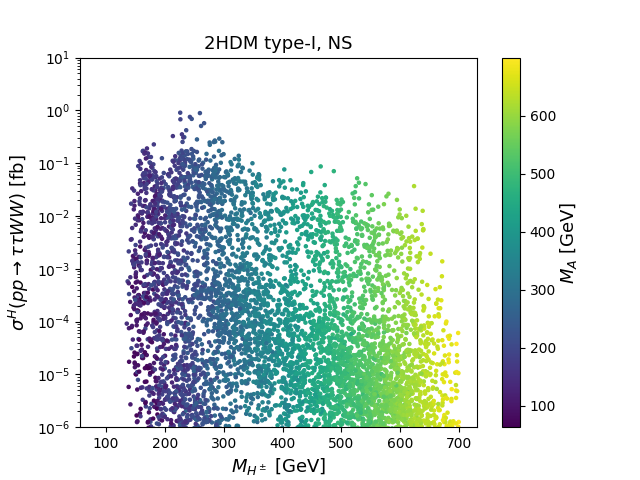} \\
		\includegraphics[width=0.46\textwidth]{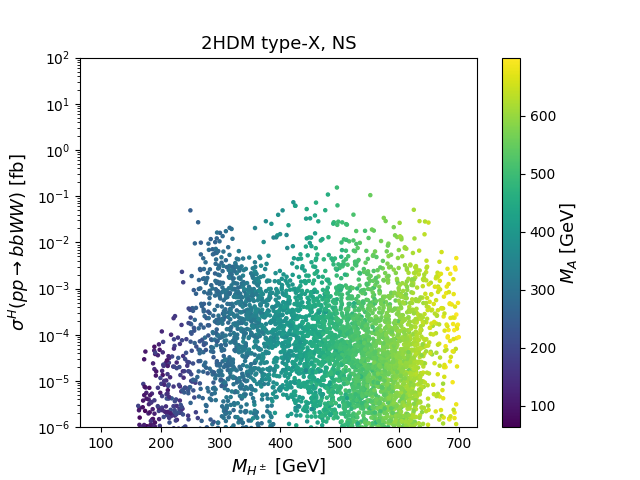} &
		\includegraphics[width=0.46\textwidth]{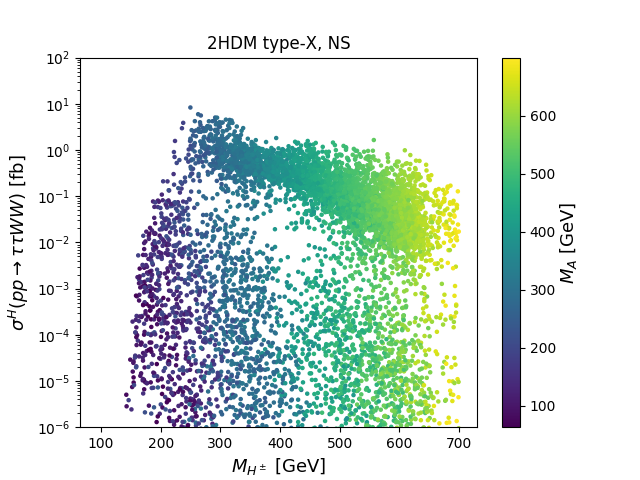} 
	\end{tabular}
	\caption{$\sigma^{H}{(pp \rightarrow bbWW)}$ (left panel), $\sigma^{H}{(pp \to \tau\tau WW)}$ (right panel) as a function of $M_{H^\pm}$, with the color code showing $M_A$. Upper (lower) panels present the type-I (type-X) results.} \label{HcWH_A:signaturetp1}
\end{figure}

\begin{table}[H]
	\centering
	\setlength{\tabcolsep}{4.3pt}
	\begin{tabular}{|c||c|c|c|c|c|c|c|}\hline
		&$M_h~[\mathrm{GeV}]$&$M_H~[\mathrm{GeV}]$&$M_A~[\mathrm{GeV}]$&	$M_{H^\pm}~[\mathrm{GeV}]$& $\sin(\beta-\alpha)$&$\tan\beta$&$m_{12}^2~[\mathrm{GeV}^2] $ \\\hline
		NS	&$125.09$&$[126;\,700]$&$[15;\,700]$&$[80;\,700]$& $[0.95;\,1]$&$[2;\,25]$&$[0;\,m_H^2\cos\beta\sin\beta]$\\\hline
		IS	&$[15;\,120]$&$125.09$&$[15;\,700]$&$[80;\,700]$& $[-0.5;\,0.5]$&$[2;\,25]$&$[0;\,m_h^2\cos\beta\sin\beta]$\\\hline
	\end{tabular}		
	\caption{2HDMs type-I and type-X input parameters.} \label{param_scans}
\end{table}
In the presentation of our results, we concentrate on the following signatures :
\begin{align}
\sigma^{S}(pp \rightarrow xWW) = \sigma(pp \rightarrow H^\pm W^\mp \rightarrow W^\pm S W^\mp \rightarrow x W^\pm W^\mp), 
\end{align}
where $S$ can be either $h$, $H$ or $A$, and $x$ can be $bb$ or $\tau\tau$. We then expect two $W$ bosons and a pair of bottom quarks or tau leptons as signatures of a charged Higgs boson. The $14$ TeV cross-section is calculated using the sophisticated public tool MadGraph@aMC\_NLO\cite{Alwall:2014hca}.\\
The light CP-even Higgs boson $h$ is believed to be the source of the reported $125$ GeV signals at the LHC in the NS. This case shows a collider phenomenology that is present over the majority of the parameter space. In the fig.\ref{HcWH_A:signaturetp1}, we show $\sigma^{H}{(pp \rightarrow bbWW)}$ (left panel), $\sigma^{H}{(pp \to \tau\tau WW)}$ (right panel) as a function of $M_{H^\pm}$, with the color code showing $M_A$. Upper (lower) panels present the type-I (type-X) results. From this figure, it’s evident that $bbWW$ is dominant in type I while $\tau\tau WW$ is promising for type X. However, these signals are negligible with respect to the expected QCD background in type I, while, the leptonic $\tau$ decay in type X enforced signal over background.

\begin{figure}[H]
	\centering
	\begin{tabular}{cc}
		\includegraphics[width=0.46\textwidth]{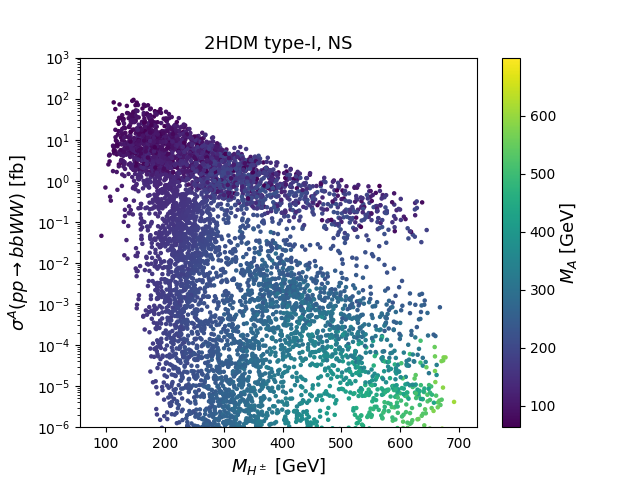} &
		\includegraphics[width=0.46\textwidth]{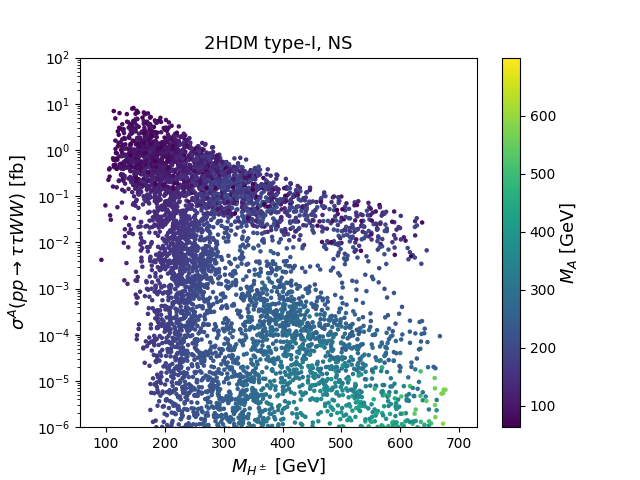} \\
		\includegraphics[width=0.46\textwidth]{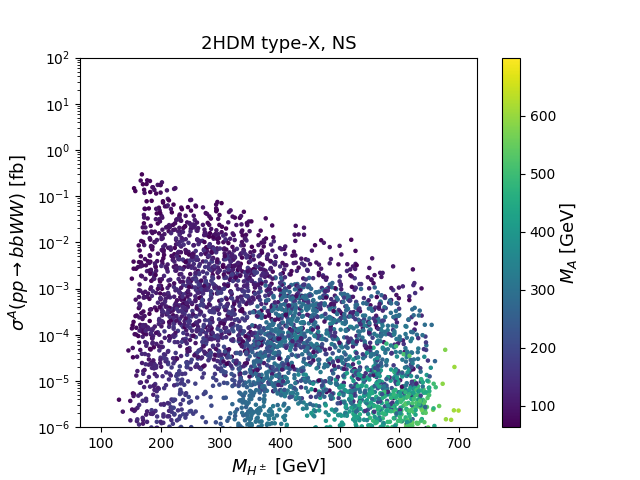} &
		\includegraphics[width=0.46\textwidth]{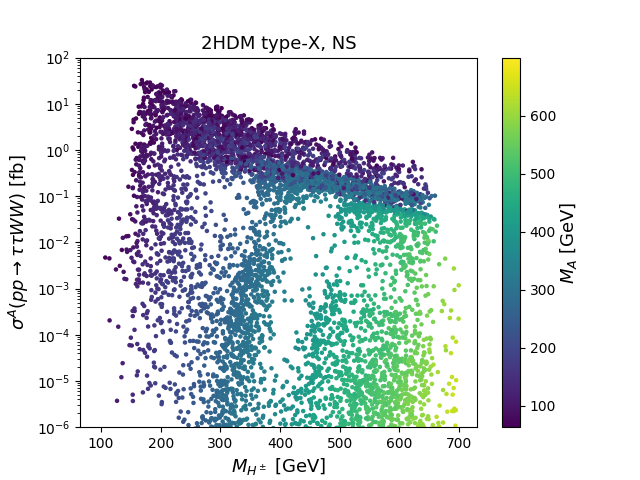} 
	\end{tabular}
	\caption{$\sigma^{A}{(pp \rightarrow bbWW)}$ (left panel), $\sigma^{A}{(pp \to \tau\tau WW)}$ (right panel) as a function of $M_{H^\pm}$, with the color code showing $M_A$. Upper (lower) panels present the type-I (type-X) results.} \label{HcWH_A:signaturetpx}
\end{figure}

In the Fig\ref{HcWH_A:signaturetpx}, we show $\sigma^{A}{(pp \rightarrow bbWW)}$ (left panel), $\sigma^{A}{(pp \to \tau\tau WW)}$ (right panel) as a function of $M_{H^\pm}$, with the color code showing $M_A$. Upper (lower) panels present the type-I (type-X) results. The signals $bbWW$ and $\tau\tau WW$ are important in this case where the charged Higgs decay following, $H^\pm \to WA$. We note that $\tau\tau$ could overwhelm the huge expected QCD background if we require at least one leptonic decay of tau leptons.

\begin{figure}[H]
	\centering
	\includegraphics[width=0.46\textwidth]{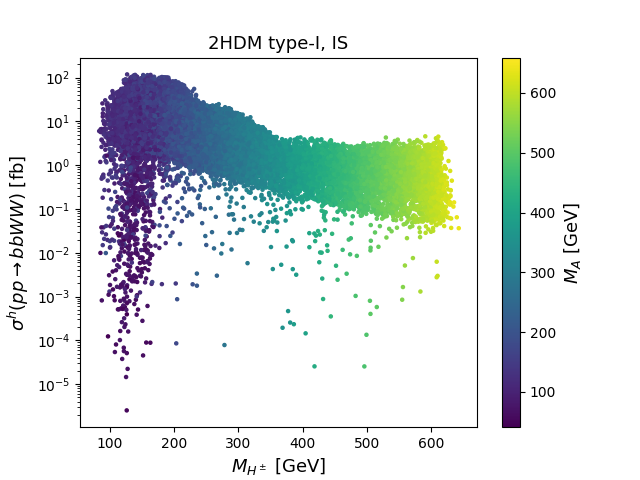} 
	\includegraphics[width=0.46\textwidth]{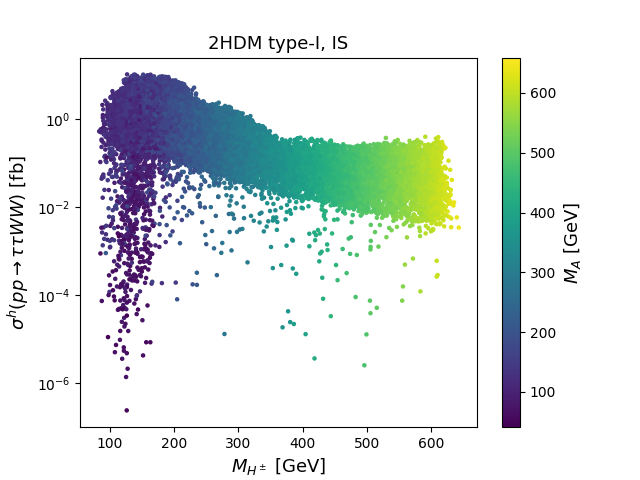} \\
	\includegraphics[width=0.46\textwidth]{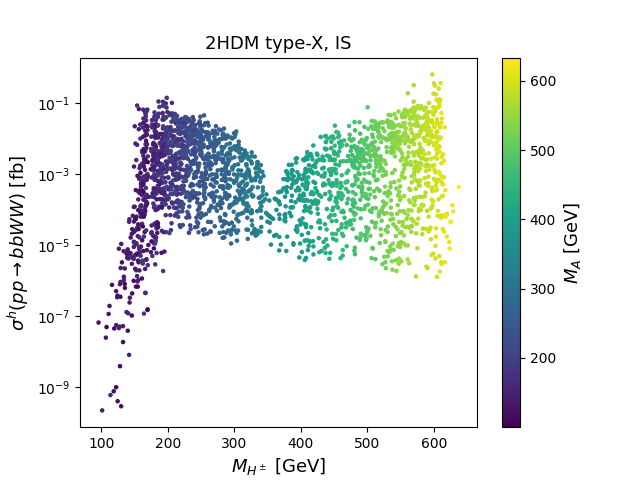} 
	\includegraphics[width=0.46\textwidth]{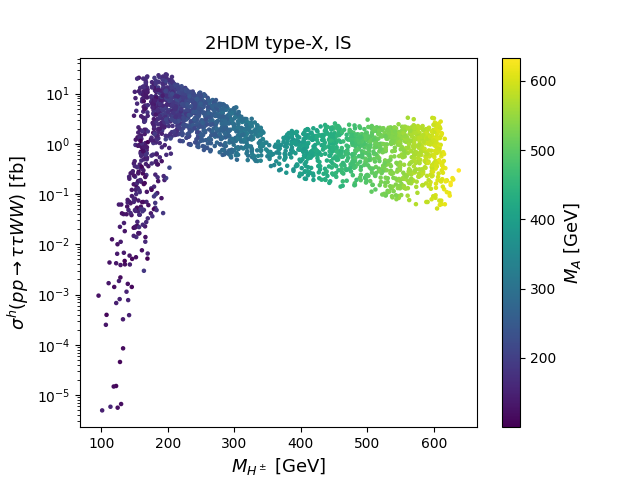} 
	\caption{$\sigma^h{(pp \rightarrow bbWW)}$ (left panel) and $\sigma^h{(pp \to \tau\tau WW)}$ (right panel) as a function of $M_{H^\pm}$, with the color code showing $M_A$. Upper (lower) panels present the type-I (type-X) results.} \label{HcW:signatures}
\end{figure}

By fixing $M_H = 125.09$ GeV, we can explain the observed Higgs signal in the Inverted scenario (IS) by the CP-even Higgs state $H$. In this realization, the $H^\pm \to W^\pm h$ decay is enhanced while $H^\pm \to W^\pm H$ channel is suppressed, compared to former scenario NS. With this configuration in mind, we show in the Fig.\ref{HcW:signatures} the signals $\sigma^h{(pp \rightarrow bbWW)}$ (left panel) and $\sigma^h{(pp \to \tau\tau WW)}$ (right panel) as a function of $M_{H^\pm}$, with the color code showing $M_A$. Upper (lower) panels present the type-I (type-X) results. In type-I, $bbWW$ and $\tau\tau WW$ reach their maximum at low $M_A$ and $M_{H^\pm}$, especially above the $H^\pm \to W^\pm A$ threshold while, in type-X, the signals $bbWW$ and $\tau\tau WW$ are interesting in both intermediate and large masses of $A$ and $H^\pm$. As expected, the $\tau\tau WW$ cross section is the dominant one in type-X instead of $b\bar{b}$ in type-I. As in the previous scenario, the background is considered a challenge in this case (IS). However, with the same choice of  $\tau$ decay in type-X, the signal could dominate the background.

\begin{figure}[H]
	\centering
	\begin{tabular}{cc}
		\includegraphics[width=0.46\textwidth]{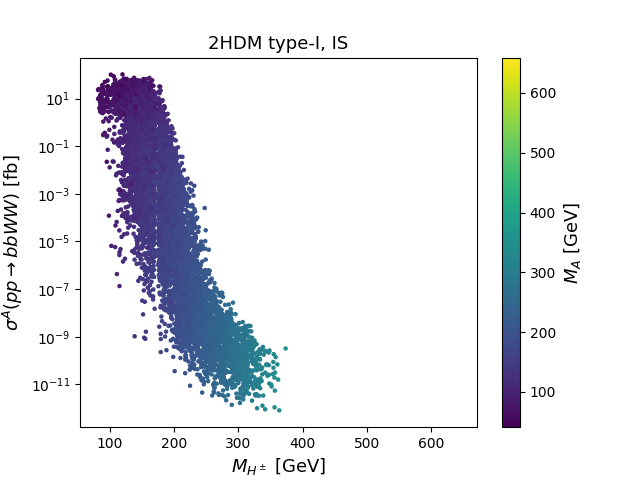} &
		\includegraphics[width=0.46\textwidth]{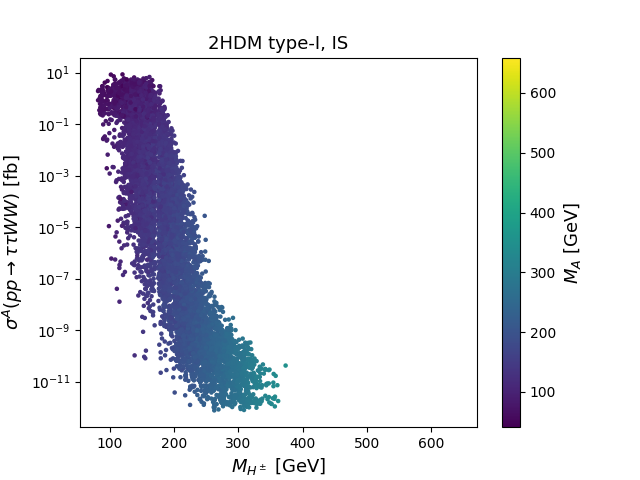} \\
	\end{tabular}
	\caption{$\sigma^A{(pp \rightarrow bbWW)}$ (left panel) and $\sigma^A{(pp \to \tau\tau WW)}$ (right panel) as a function of $M_{H^\pm}$, with the color code showing $M_A$.} \label{HcW-signatures22}
\end{figure}

We now investigate the region below the $H^\pm \to W^\pm A$ decay threshold, where the latter is dominant. Figure \ref{HcW-signatures22} depicts the same ultimate states as Figure \ref{HcW:signatures}, with the exception of $H^\pm \to W^\pm A$ instead of $H^\pm \to W^\pm h$ and just in type-I only. As it's clear from this figure, the such signatures are important in the small masses of $A$ and $H^\pm$. In the type-X, these signatures are negligible since the $\tau\nu_\tau$ decay dominates at small values of $M_{H^\pm}$ and $M_A$.

\section{Benchmarks Points (BPs)}
In Table\ref{Table:BPs}, we suggest six benchmark points for 2HDMs type-I in both normal and inverted Scenario, with the purpose of facilitating the search for $pp\to H^\pm W^\mp$ signatures at the LHC.
\label{sec:BPs}
{\renewcommand{\arraystretch}{0.2} 
	{\setlength{\tabcolsep}{0.3cm} 
	\begin{table}[H]
		\centering
		\begin{tabular}{|c|c|c|c|c|c|c|}\hline	
			Parameters & BP1& BP2 & BP3 & BP4 & BP5 & BP6  \\\hline
			\multicolumn{7}{|c|}{\small Normal scenario(NS)}\\\hline				
			$M_h$ (GeV) & $125.09$ & $125.09$ & $125.09$ & $125.09$ & $125.09$ & $125.09$ \\
			$M_H$ (GeV) & $135.07$ & $144.62$ & $132.07$ & $130.26$ & $135.12$ & $134.75$\\
			$M_A$ (GeV)& $200.95$ & $219.65$ & $67.01$ & $74.07$ & $62.97$ &$66.54$\\  
			$M_{H^\pm}$ (GeV) & $226.20$ & $259.66$ & $146.88$ & $144.66$ & $113.18$ &$123.09$\\ 
			$\sin(\beta-\alpha)$ & $0.994$ & $0.985$  & $0.989$ & $0.985$ & $0.991$ & $0.968$\\ 
			$\tan\beta$ & $3.97$ & $2.77$ & $3.53$ & $3.55$ & $4.26$& $4.37$ \\
			$m_{12}^2$ (GeV$^2$) & $4322.16$ & $6675.8$ & $4565.08$ & $4417.98$ & $4055.61$ & $3949.10$\\ \hline
			\multicolumn{7}{|c|}{\small BR($H^\pm \to XY$) in \%}\\\hline			
			BR($H^\pm \to W^\pm H$) & $35.41$ & $46.78$ & $-$ & $-$ & $-$ & $-$\\
			BR($H^\pm \to W^\pm A$) & $-$ & $-$ & $98.12$ & $95.37$ & $92.47$ & $95.19$\\\hline
			\multicolumn{7}{|c|}{\small BR($h \to XY$) in \%}\\\hline				    
			BR($H \to b\bar{b}$) & $53.68$ & $29.76$ & $11.01$ & $11.20$ & $1.68$ & $0.16$ \\
			BR($H \to \tau \tau$) & $5.26$ & $2.95$ & $1.07$ & $1.09$ & $0.16$& $0.01$\\
			BR($H \to \gamma\gamma$) & $0.34$ & $0.33$ & $0.11$ & $0.24$ & $0.03$ & $0.07$\\ \hline
			\multicolumn{7}{|c|}{\small BR($A \to XY$) in \%}\\\hline			
			BR($A \to b\bar{b}$) & $22.54$ & $17.96$ & $79.98$ & $78.51$ & $80.80$ & $80.08$ \\
			BR($A \to \tau \tau$) & $2.43$ & $1.97$ & $6.96$ & $6.97$ & $6.94$ & $6.95$  \\
			BR($A \to \gamma\gamma$) & $0.05$ & $0.05$ & $0.01$ & $0.01$ & $0.01$ &$0.01$ \\\hline
			\multicolumn{7}{|c|}{\small $\sigma$ in fb}\\\hline			
			$\sigma^H(pp \to bbWW)$ & $9.23$ & $8.95$ & $-$ & $-$ & $-$ & $-$ \\
			$\sigma^H(pp \to \tau\tau WW)$ & $0.90$ & $0.88$ & $-$ & $-$ & $-$ & $-$\\
			$\sigma^H(pp \to \gamma\gamma WW)$ & $0.05$ & $0.09$ & $-$ & $-$ & $-$ & $-$\\\hline
			$\sigma^A(pp \to bbWW)$ & $-$ & $-$ & $93.43$ & $88.32$ & $81.14$ & $72.36$\\
			$\sigma^A(pp \to \tau\tau WW)$ & $-$ & $-$ & $8.13$ & $7.84$ & $6.97$ & $6.28$ \\
			$\sigma^A(pp \to \gamma\gamma WW)$ & $-$ & $-$ & $0.02$ & $0.02$ & $0.01$ & $0.01$\\ \hline
			\multicolumn{7}{|c|}{\small Inverted scenario (IS)}\\\hline
			$M_h$ (GeV) & $64.68$ & $68.22$ & $69.29$ & $112.45$ & $115.42$ & $71.68$ \\
			$M_H$ (GeV) &  $125.09$ & $125.09$ & $125.09$  & $125.09$  & $125.09$ & $125.09$ \\
			$M_A$ (GeV)&  $130.84$ & $147.98$ & $132.88$  & $53.72$  & $51.90$  & $135.54$ \\  
			$M_{H^\pm}$ (GeV) & $126.68$ & $139.15$ & $163.20$  & $101.36$  & $119.45$ & $115.38$ \\ 
			$\sin(\beta-\alpha)$ & $0.127$ & $0.140$ & $-0.062$ & $0.175$  & $0.134$ & $-0.144$ \\ 
			$\tan\beta$ & $3.46$ & $3.35$ & $3.13$ & $4.02$ & $3.80$ & $6.94$ \\
			$m_{12}^2$ (GeV$^2$) & $1053.71$ & $511.93$ & $850.14$ & $2757.59$  & $2782.55$ & $177.81$ \\ \hline
			\multicolumn{7}{|c|}{\small BR($H^\pm \to XY$) in \%}\\\hline			
			BR($H^\pm \to W^\pm h$) & $94.72$ & $95.95$ & $99.54$ & $-$ & $-$ & $94.11$ \\
			BR($H^\pm \to W^\pm A$) & $-$ & $-$ & $0.03$ & $90.00$ & $97.52$ & $-$ \\\hline
			\multicolumn{7}{|c|}{\small BR($h \to XY$) in \%}\\\hline				    
			BR($h \to b\bar{b}$) & $85.76$ & $85.49$ & $85.39$ & $5.38$ & $1.08$ & $9.71$ \\
			BR($h \to \tau \tau$) & $7.37$ & $7.41$ & $7.43$ & $0.51$ & $0.10$ & $0.85$ \\
			BR($h \to \gamma\gamma$) & $<0.01$ & $<0.01$ & $0.02$ & $< 0.01$ & $< 0.01$ & $51.14$ \\ \hline
			\multicolumn{7}{|c|}{\small BR($A \to XY$) in \%}\\\hline			
			BR($A \to b\bar{b}$) & $30.88$ & $16.29$ & $36.79$ & $82.60$ & $82.94$ & $13.45$ \\
			BR($A \to \tau \tau$) & $3.07$ & $1.66$ & $3.67$ & $6.87$ & $6.85$ & $1.35$ \\
			BR($A \to \gamma\gamma$) & $0.02$ & $0.02$ & $0.03$ & $0.01$ & $0.01$ & $0.01$ \\\hline
			\multicolumn{7}{|c|}{\small $\sigma$ in fb}\\\hline			
			$\sigma^h(pp \to bbWW)$ & $118.15$ & $115.14$ & $115.07$ & $-$  & $-$ & $3.65$ \\
			$\sigma^h(pp \to \tau\tau WW)$ & $10.16$ & $9.99$ & $10.01$ & $-$  & $-$ & $0.32$\\
			$\sigma^h(pp \to \gamma\gamma WW)$ & $<0.01$ & $<0.01$ & $0.03$ & $-$  & $-$ & $19.24$\\\hline
			$\sigma^A(pp \to bbWW)$ & $-$ & $-$ & $0.02$ & $100.57$  & $103.02$ & $-$ \\
			$\sigma^A(pp \to \tau\tau WW)$ & $-$ & $-$ & $<0.01$ & $8.37$  & $8.51$ & $-$\\
			$\sigma^A(pp \to \gamma\gamma WW)$ & $-$ & $-$ & $ < 0.01$ & $0.01$  & $0.01$ & $-$\\ 
			\hline					 				  
		\end{tabular}
		\caption{2HDMs type-I selected BPs in Both Scenarios;  NS and IS.}
		\label{Table:BPs}
	\end{table}
\section{Conclusion}
\label{sec:conclusion}
Based on a 2HDMs type-I and X framework, we examined the production of a single-charged Higgs boson at the LHC with 14 TeV using the $pp\to H^\pm W^\mp$ processes, taking into account the most recent experimental and theoretical constraints. we focus on the $H^\pm \to W^\pm hi/A$ pathway and we have searched for $WW$ associated with $bb$ and $\tau\tau$ final states. We not that the $bbWW$ signal is contaminated by the huge QCD background from $t\bar{t}$, while the $\tau\tau WW$ can give the best signal since they could suppress that background using $\tau$ leptonic decay and experimental techniques. 
\bibliographystyle{JHEP}
\bibliography{Bib_ref}
\end{document}